# Quantum Walks in Periodic and Quasiperiodic Fibonacci Fibers

Dan T. Nguyen, Thien An Nguyen, Rostislav Khrapko, Daniel A. Nolan and Nicholas F. Borrelli

*Science and Technology Division, Corning Research and Development Corporation, Sullivan Park, Corning, NY 14831*

*Email: nguyendt2@corning.com*

**Abstract**

Quantum walk is a key operation in quantum computing, simulation, communication and information. Here, we report for the first time the demonstration of quantum walks and localized quantum walks in a new type of optical fibers having a ring of cores constructed with both periodic and quasiperiodic Fibonacci sequences, respectively. Good agreement between theoretical and experimental results have been achieved. The new multicore ring fibers provide a new platform for experiments of quantum effects in low-loss optical fibers which is critical for scalability of real applications with large-size problems. Furthermore, our new quasiperiodic Fibonacci multicore ring fibers provide a new class of quasiperiodic photonics lattices possessing both on- and off-diagonal deterministic disorders for realizing localized quantum walks deterministically. The proposed Fibonacci fibers are simple and straightforward to fabricate and have a rich set of properties that are of potential use for quantum applications. Our simulation and experimental results show that, in contrast with randomly disordered structures, localized quantum walks in new proposed quasiperiodic photonics lattices are highly controllable due to the deterministic disordered nature of quasiperiodic systems.

# 1. Introduction

For the last decade there has been significant of efforts in the investigation of quantum walks (QWs) in photonics lattices or arrays of waveguides [1-5]. QWs have emerged from fundamental research to be one of the key operations in quantum computing, simulation, communication and information. Quantum walks have been modeled for exponential speedup in quantum algorithms [6, 7], to implement universal quantum gates for quantum computers [8, 9] and quantum simulations [10, 11] etc. These are among the most promising schemes for their unprecedented computing acceleration that can solve classically intractable problems. The first scalable quantum simulations of molecular energies have been demonstrated a few years ago [12-14] indicating that quantum revolution 2.0 may be viable not very far away. Photons with their dual wave-particle nature have been demonstrated as excellent "walkers" since they are easily generated and manipulated. Furthermore, photonics technology is very mature, and many photonics systems can be operated ideally in room-temperature conditions. That is a huge advantage of photonics as compared to other technologies used in quantum applications such as ion-trapped, superconducting operating at extremely low temperatures, often at tens of millikelvin (mK). In light of that direction, quantum simulations with light [15] and photonics quantum gates [16] have been proposed and developed recently. Beam splitter arrays have been used to perform discrete-time QWs (DTQW) [17, 18], and evanescently coupled parallel arrays of waveguides have been proven to be excellent platform to perform continuous-time QWs (CTQW) [1-5].

One unique phenomenon in QWs is localization in the presence of a disordered medium. This phenomenon, commonly known as Anderson localization, is usually discussed in terms of coherent evolution (e.g. the quantum walks) in the presence of a disordered medium. By breaking the periodicity of the evolution through spatial and/or temporal randomizing the operations with which the dynamics of the system is determined, the effect of a randomly disordered medium can be simulated, and localizations has been realized [19 - 21]. Since then, there have been increased interests and efforts on investigation not only of fundamentals but also applications of LQWs, partially motivated by the possibility of employing localized photonic states for a secure quantum memory [22] and secure transmission of quantum information [23]. Anderson localization arises from destructive interferences among different scattering paths of a quantum particle propagating in a static disordered medium [24]. Sixty years after its discovery, Anderson localization is still widely studied in many different areas of physics [25]. For the last few years, localization of light in a randomly disordered medium has become new research field for discovering many useful effects that are significant for many applications [26-29]. Meanwhile, it is well known that deviations from periodicity may result in higher complexity and give rise to a number of surprising effects. One such deviation can be found in the field of optics in the realization of photonic quasi-crystals, a class of structures made from building blocks that are arranged using well-designed patterns but lack translational symmetry. A quasiperiodic system is neither a periodic nor a random one so it could be considered as an intermediate between the two. It has been recognized that quasiperiodic systems could also lead to localization in optics [30 - 32]. Examples of such systems constructed with Fibonacci sequences include 1D quasi-crystalline Fibonacci dielectric multilayers (FDML) [33] and semiconductor quantum-wells [34], two-dimensional (2D) quasi-crystalline structures [35], and three-dimensional (3D) quasi-crystals [36].

It is worth to note that, disorder-induced localization can be conventionally quantified by averaging over realizations on many systems having the same degree of disorder. Similarly, LQWs in integrated photonics systems have been realized on many randomly disordered arrays of evanescently coupled waveguides. The final results are averaged over all realizations in such arrays of waveguides whose randomness is controlled within a defined range of the disorder. Because of that, realization of LQWs has been proven experimentally to be difficult [19]. Meanwhile, quasiperiodic systems or quasi-crystals provides deterministic disorder deviated from periodicity resulting in localization of light deterministically in those systems [30-33], meaning there is no need to do averaging over many samples or systems. In that spirit, we have proposed theoretically a new class of quasiperiodic photonics lattices (QPLs) constructed with Fibonacci and Thue-Morse sequences that can be used to realize deterministically LQWs [37, 38]. Furthermore, our simple construction rules allow us to create symmetrically quasi-periodic QPLs. Consequently, LQWs with symmetrical probability distributions can be realized deterministically in the QPL structures.

QWs of photons have been realized in several integrated optical platforms such as silicon oxynitride waveguides [1], laser-writing waveguides in silicon-on-insulator (SOI) [2, 3], in silica glass waveguides [4], and in borosilicate glass waveguides [5]. All those configurations have a common feature that the 'walking region' are the waveguide arrays. In such structures, the spacing between waveguides in an array that is required for evanescent coupling (on the order of several micrometers). It is well-known that, couplings between silicon-based waveguides (mode field diameter MFD ~ 1 micron) with single-mode (SM) optical fiber (typically MFD ~ 10 micron) is usually of very high loss (more than 20 dB) and require significant efforts. Moreover, intrinsic background losses in silicon waveguides (0.1 dB/cm to ~1 dB/cm) are few orders of magnitudes higher than that of optical fibers (~ 0.2 dB/km) for telecom signals. High insertion loss including intrinsic and coupling losses could be a very negative factor for scaling up numbers of waveguides in silicon-photonics lattices for realizing QWs. Meanwhile, laser-writing technique can write large number of waveguides on glass, for example a lattice of 49x49 = 2401 waveguides in [5]. However, laser-written glass waveguides usually have small index contrasts, resulting in weak confinement and sensitivity to the imperfections and perturbations as clearly indicated in recent experiments in which QWs characteristics were completely distorted in just about 1cm of propagation [5]. The optical fiber-based QWs systems such as multicore fibers would significantly reduce or even eliminate these problems. The multicore fibers with extremely low insertion loss would be critical for scalable platforms for QWs, and quantum photonics computing and simulations to solve large-size problems which are intractable on classical computing.

In this paper, we have demonstrated for the first time QWs and LQWs in a new type of optical fibers having cores constructed with periodic and quasiperiodic Fibonacci sequence, respectively. Good agreement between theoretical and experimental results have been achieved. The new multicore-ring fibers (MCRF) provide a new scalable platform for experiments of QWs and other quantum effects that require large numbers of waveguides. More important, our new Fibonacci MCRF (FMCRF) provide a new class of photonics lattices possessing both *on- and off-diagonal deterministic disorder* for realizing LQWs deterministically. The proposed FMCRFs are simple and straightforward to make and have a rich set of properties that are of potential use for quantum applications. Furthermore, our simulation and experiment results show that, in contrast with randomly disordered structures, LQWs in quasi-periodic ones are highly controllable due to the deterministic disordered nature of quasi-periodic systems.

**2. New Periodic and Quasiperiodic Multicore Ring Fibres**

In this section, we will present in details the concept of new multicore ring fibers as platform for realizing QWs and LQWs. In general, our new type of fibers - the multicore ring fibers (MCRF) are designed and fabricated with a configuration of a ring of cores and is compact in comparison with planar arrays of waveguides used to realize QWs on a line [1-4]. Fig. 1 shows the diagram of MCRF and the image of the fabricated fiber of 39 cores and the image of fiber that is illuminated and overfilled with light of wavelength 1.55 μm. All the cores of the fiber have diameter $a = 4.5$ μm, $R_1 = 120$ μm, and $R_2 = 160$ μm.

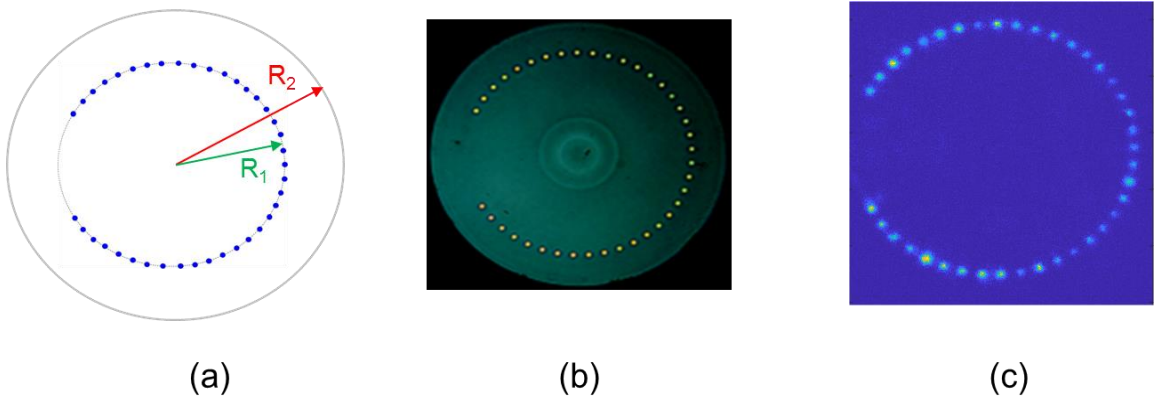

(a) (b) (c)

**Figure 1**. Diagram of MCRF with 39 identical SM cores (a), image of the fabricated fiber (b), and the image of the fiber that is overfilled with light of wavelength 1.550 μm.

The fiber shown in Fig. 1 is designed and fabricated with all cores of identical SM waveguides that are regularly *periodic* in a circular ring. The image of the overfilled fiber shows there is variation of the cores, and our characterizations show core diameter variation of about +/- 2%. This periodic multicore ring fiber is called multicore-ring fiber (MCRF) for short. In our experiments of single-photon QWs, we will launch signal into the center core of the core ring, and the QWs process will take place from the center to the end cores of the two symmetrical arms. Notice that, the two end-cores of the two arms should not be too close to avoid coupling between these two cores that would distort distribution of the QWs on a line, which do not have such coupling in a planar array of waveguides [1-4].

For the quasi-periodic multi-core ring fibers, the ring of cores is constructed with a Fibonacci sequence with two different SM waveguides *A* and *B* (the fiber is called Fibonacci MCRF or FMCRF). For example, waveguide $A$ and $B$ is characterized by $V_A = \pi a_A NA_A/\lambda$, and $V_B = \pi a_B NA_B/\lambda$, respectively, where $a_{A(B)}$ stands for core diameter and $NA_{A(B)}$ is the numerical aperture of waveguide $A(B)$. Note that, the numerical aperture NA can be determined by the index difference between core and clad $\Delta n = n_{core} - n_{clad}$, and we will use $\Delta n$ to characterize waveguides in our calculations. In general, the construction rule for the cores of a fiber with Fibonacci j-th order are the same as in the Fibonacci arrays of waveguides or photonics lattices [37, 38]. The difference between the two structures is the ring of cores in FMCRF instead of linear arrays of cores in Fibonacci array of waveguides. The construction rule for the cores of Fibonacci jth order is defined as

$$F_j = S_j S_{j-1} \cdots S_2 S_1 S_2 \cdots S_{j-1} S_j, \tag{1}$$

where $S_1, S_2 \ldots S_j$ are Fibonacci elements defined as

$$S_j = S_{j-2} S_{j-1}, \;\; with \;\; S_1 = A, \;\; S_2 = B. \tag{2}$$

Here, $A$ and $B$ are two different SM waveguides as described above (see more details of the Fibonacci quasiperiodic photonics lattice in our previous works [37, 38]).

As examples, Fig. 2 shows diagrams of structures of core rings of $4^{th}$, $5^{th}$ and $6^{th}$ order of FMCRFs. Notice that, these three core rings have the same ring radius for convenient discussion but not necessary. The red arrows indicate the input core, which is the center of the ring composed of two symmetrical arms. $S_1, S_2 \ldots S_j$ are the Fibonacci elements of $j^{th}$ order defined in Eq. (2) above and are indicated by color circles. From the definition in Eq. (1), it is easy to calculate the number of cores for the $4^{th}$, $5^{th}$ and $6^{th}$ order FMCRFs are 13, 23 and 39, respectively. It is clear from Fig. 2 that the core rings constructed with Fibonacci sequences of two different SM waveguides $A$ and $B$ are *on-diagonal* quasiperiodic due to the Fibonacci distributions of propagation constants $\beta_A$ and $\beta_B$ (see some description of on- and off-diagonal disordered arrays of waveguides in [21]). Notice that, the coupling coefficients between the nearest waveguides are the functions of the overlapping between the modes and the propagation constants of these waveguides [37]. Consequently, the coupling coefficients in our new FMCRFs also have quasiperiodic – or deterministically disordered distribution. Therefore, our FMCRFs provide platforms having *both on- and off-diagonal* deterministic disorders for realizing LQWs deterministically (e.g., both propagation constants and coupling coefficients are quasiperiodic).

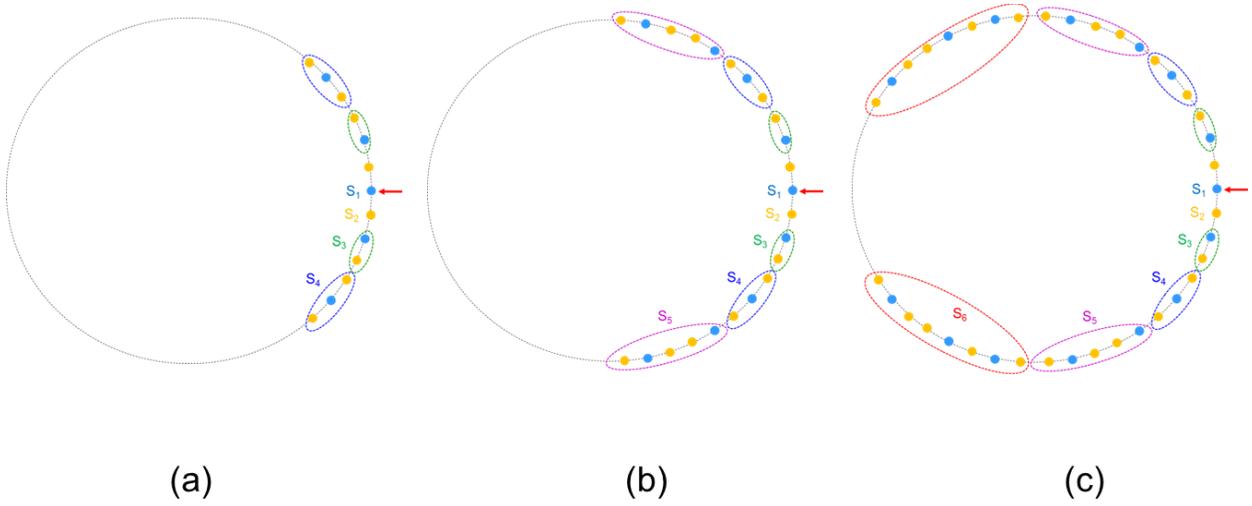

**Figure 2**. Diagram of ring of cores: 4$^{th}$ order FCRF4 with 13 cores (a), 5$^{th}$ order FCRF5 with 23 cores (b), and 6$^{th}$ order FCRF6 fiber with 39 cores (c). Red arrow indicates input core.

## 3. Quantum Walks in Periodic and Quasiperiodic Multicore Ring Fibers

In this section, first we will present the simulation of QWs and LQWs in MCRF and FMCRF. In general, simulations of single photon QWs in *irregular* arrays of waveguides are extremely difficult and therefore numerical solutions are necessary. Especially, if the propagation loss is included in numerically solving the Linblad equation it would require large resources including computing time and memory [40, 41]. However, we can make use of the fact that single photon

QWs do not exhibit any different behavior from classical wave propagation, and the light intensity distribution corresponds to the probability of detecting the photon at any position [42]. Therefore, we can use the beam propagation method (BPM) [43] to simulation of single photon QWs in both periodic MCRF and quasiperiodic FMCRF. The BPM can describe very well the evanescent coupling between waveguides, and the method only requires fundamental parameters of the system such as core sizes, indexes and positions of each waveguide [44]. We have our own Matlab codes to simulate beam propagations in complicated structures of waveguides. The method has been successfully applied to simulate and design Yb-doped multicore fiber lasers for the coherent Ising machine [45, 46], and also for single-photon quantum QWs in regular and irregular arrays of waveguides [37, 38]. Details of the BPM method are presented generally in [43, 44], its applications for simulating QWs in photonics lattices [38, 39] and multicore fiber lasers in [45, 46], and the simulation results of QWs in MCRF and FMCRF are presented as follows.

The simulations in Fig. 3 show the probability distribution of QWs in periodic MCRF with photons spread across the lattice by coupling from one waveguide to its neighbors in a pattern characterized by two strong ''ballistic'' lobes as in a typically normal QWs on a line. Meanwhile, the results for FMCRFs are structurally different: LQWs are clearly shown in quasiperiodic Fibonacci multicore fibers FMCRFs. Furthermore, symmetrical distributions of LQWs in FMCRFs can be achieved due to the symmetry of the quasi-periodic ring of cores in FMCRFs.

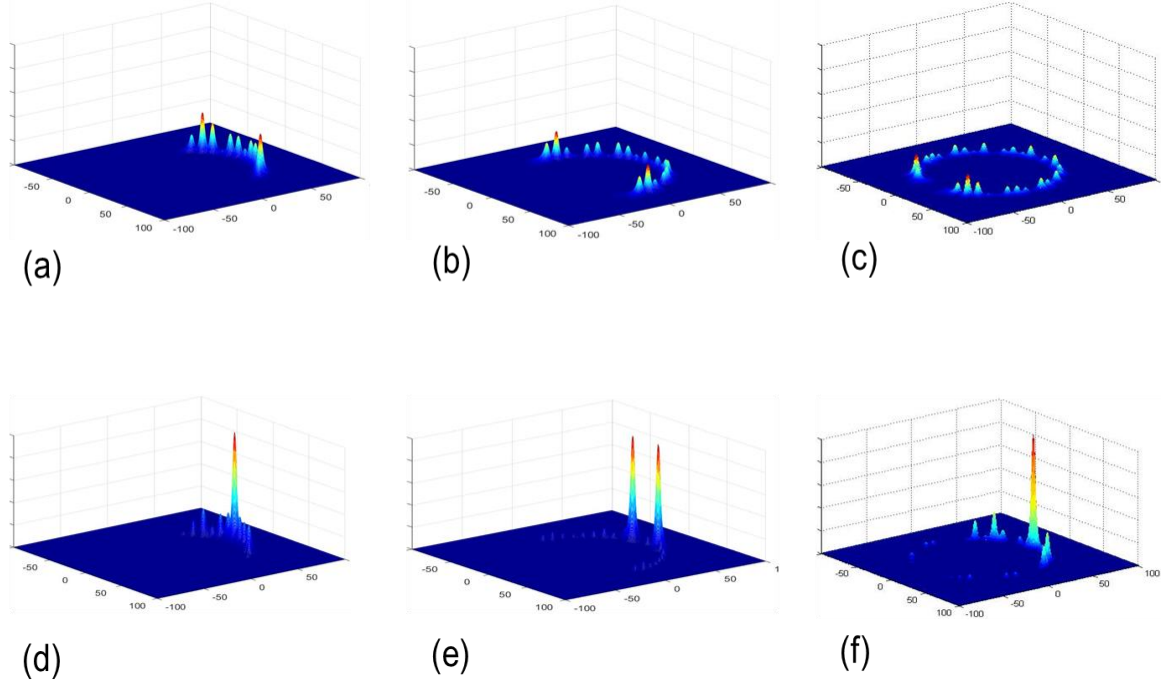

**Figure 3**. Probability distribution of photons in quantum walks: a) regular MCRF with 13 cores, b) regular MCRF with 23 cores and c) regular MCRF with 39 cores, d) FMCRF4 with 13 cores, e) FMCRF5 with 23 cores and f) FMCRF6 with 39 cores. Signal wavelength $\lambda = 1.550$ μm.

In order to demonstrate QWs and LQWs in periodic and quasiperiodic Fibonacci MCRFs (or FMCRFs), we have designed and fabricated this new type of fiber with ring of cores, both periodic and quasiperiodic ring of cores. Images of MCRF and FMCRF are shown in Figs. 5a and 6a, respectively. We have performed experiments to realize single photon QWs in those fibers. Photon distribution of QWs in the MCRF and FMCRF are presented in Figs. 5c and 6c to compare with simulation results 5b and 6b, respectively.

The two fabricated fibers, one is a periodic core ring - MCRF with 39 identical SM waveguides having core diameter $a = 4.5$ μm, index difference $\Delta n = n_{core} - n_{clad} = 0.0035$; core-ring radius $R_1 = 120$ μm, and clad diameter $R_2 = 160$ μm. The quasi-periodic FMCRF is $6^{th}$ Fibonacci order that has $R_1$ and $R_2$ are the same as in MCRF, but its 39 cores are composed of two different SM waveguides *A* and *B* with the same core-diameter $a = 4.5$ μm, and different index differences $\Delta n_A = 0.0045$ and $\Delta n_B = 0.0035$. Note that, *A*- and *B*-waveguides are depicted by blue and yellow solid circles in Fig. 2 above. Our characterizations show the variation of core diameters in FMCRF is about +/- 3% which is larger than +/-2% in MCRF which can be attributed to the different materials used in cores *A* and *B*-waveguides of FMCRF.

The MCRF and FMCRF both with 39 single mode cores were characterized using a cross-polarization microscopy. The microscopy system is a Nikon, high magnification optical microscope with error of ±0.05 μm. The average core diameter measured is ~4.40 μm and 4.55μm for MCRF and FMCRF, respectively. The average distance from center-to-center of neighboring cores is ~16.89 μm and ~16.80 μm for MCRF and FMCRF, respectively. The index observed from crossed-polarization indicate that the CRF have identical refractive index for all of cores, whereas the FCRF have different core refractive indices grouped as described in previous sections. The core ring radius is approximately ~120 μm and fiber radius is approximately ~158 μm.

Demonstration of quantum walks in MCRF and FMCRF were conducted with a stripped fiber, at approximately 4-cm. The fiber is placed on a v-groove in an imaging system shown in figure 7. A tunable source from 1510 – 1590 nm laser illuminates the MCRF/FMCRF. The steps taken to identify the central core and the measured quantum walk distribution is as follows: First, we Illuminate subsections of the fiber of interest (FOI) such that cores are illuminated. As an example, the image of the illuminated MCRF is shown in Fig. 1c. Next, we combine illumination images to identify the position of each cores using the Matlab/Labview algorithms to determine the diameter and position of each core of the FOI. From that characterization the position of the central core is determined for the input signals in the quantum walks experiments. Once we know the position of the input core, we launch a coherent beam of light into it by butt-coupling with a SM fiber that is mode-matched to the input core of MCRF and FMCRF. The image of output facet is captured and output signal from all cores are measured. We repeat those steps for wavelength sweep from 1530 – 1559nm to account for fiber length variations. Matlab codes are written to calculate the total intensity for each core.

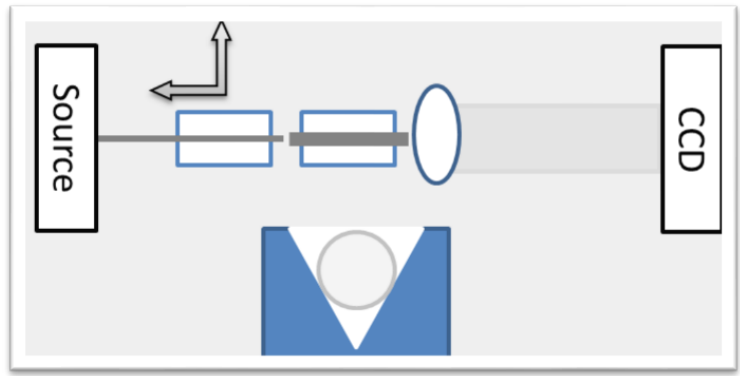

**Figure 4.** drawing of experimental set up.

We show in Figs. 5 and 6 both simulation and experimental results of single photons QWs in MCRF and FMCRF. Figs 5a and 6a are the schematic designs and Figs. 5b and 6b are images of cross section of MCRF and FMCRF, respectively. Figs 5c and 6c are the simulations and Figs 5d and 6d are experiment results of photon distribution of QWs in MCRF and FMCRF, respectively.

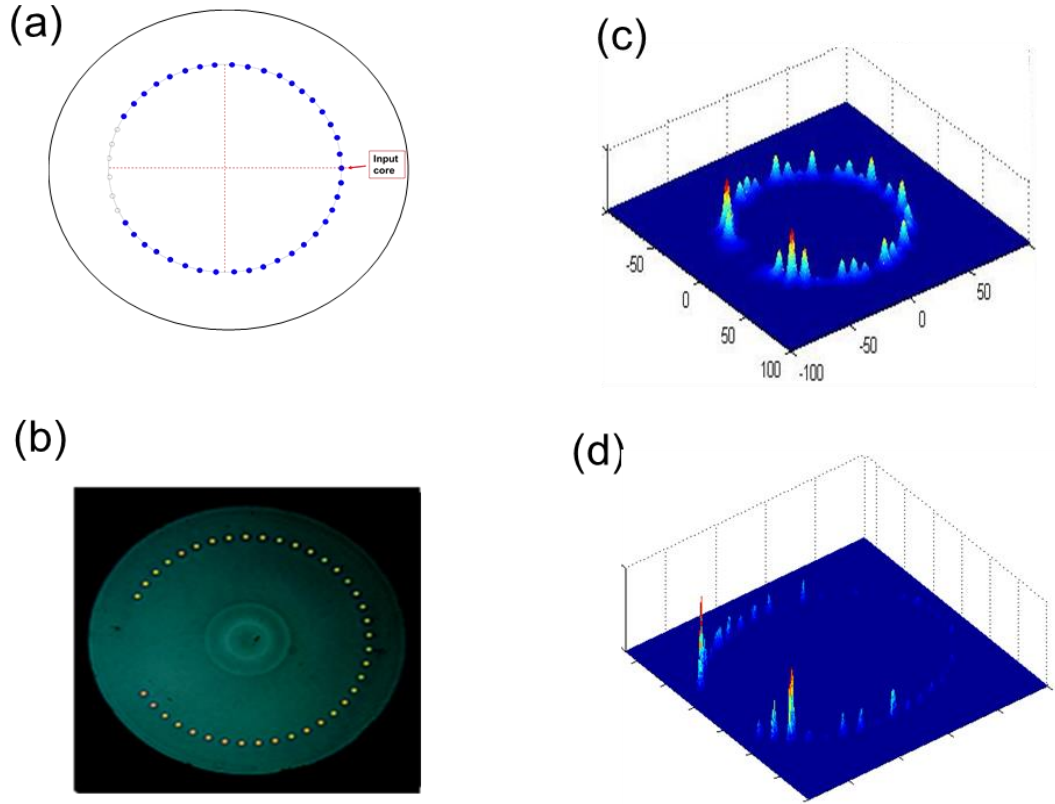

**Figure 5**. (a) Schematic of multicore fibers MCRF of 39 cores that are regular-spacing in a ring of radius *R*, all cores have same index difference $\Delta n$ and core size *a* and are single mode, (b) Image of cross section of MCRF, (c) calculated probability photon distribution of quantum walks in the MCRF, and (d) Experiment data of photon distribution. The experimental measurement of the photon distribution at around 4.1cm of MCRF and simulation results at 4.1cm are in good agreement. Especially, the typical feature of QWs with two strong lobes at the end of walking length are clearly shown in (c) and (d).

The results in Fig. 5 show probability distributions of single photons QWs in periodic MCRF spread across the lattice by coupling from one waveguide to its neighbors in a pattern characterized by two strong ''ballistic'' lobes. Note that the results are measured at ~ 4.1 cm of fiber length, and typical characteristics of QWs on a line are clearly shown. The experimental data is in very good agreement with simulation results, both qualitative and quantitative. It is worth to mention here that experiments of QWs in [3] with lattice of identical waveguides fabricated on an AlGaAs substrate waveguide is about in 8 mm long, in laser-writing photonics lattices [5] is less than 1cm. The QWs behavior can only be preserved at around 9.81 mm of the laser-written waveguides lattice, and it is completely distorted if walking longer distances. Our experimental results show QWs characteristics are clearly preserved in ~ 4cm length of MCRF indicate that both losses and imperfection in our MCRF are very good for QWs experiments.

In Figs. 6 we show the results of QWs in FMCRF at ~ 4.15cm. As stated earlier, this quasiperiodic fiber – the FMCRF is composed of two different waveguides of different materials. As a result, the FMCRF has variation of core diameters +/- 3% which is larger than in MCRF. Even with that, the strong localization in the center core as predicted in the simulation is clearly preserved in Fig. 6d. However, the imperfection of the core sizes and core-to-core distances is the reason for slight distortion of QWs behavior resulting in an unsymmetrical distribution of photons observed in the experiment. Meanwhile, modeling results in the ideal conditions predict a symmetrical distribution of QWs as shown in Fig. 6c. It is clear that the experimental results are in a very good agreement with the simulations, except for small discrepancies due to un-avoidable imperfections.

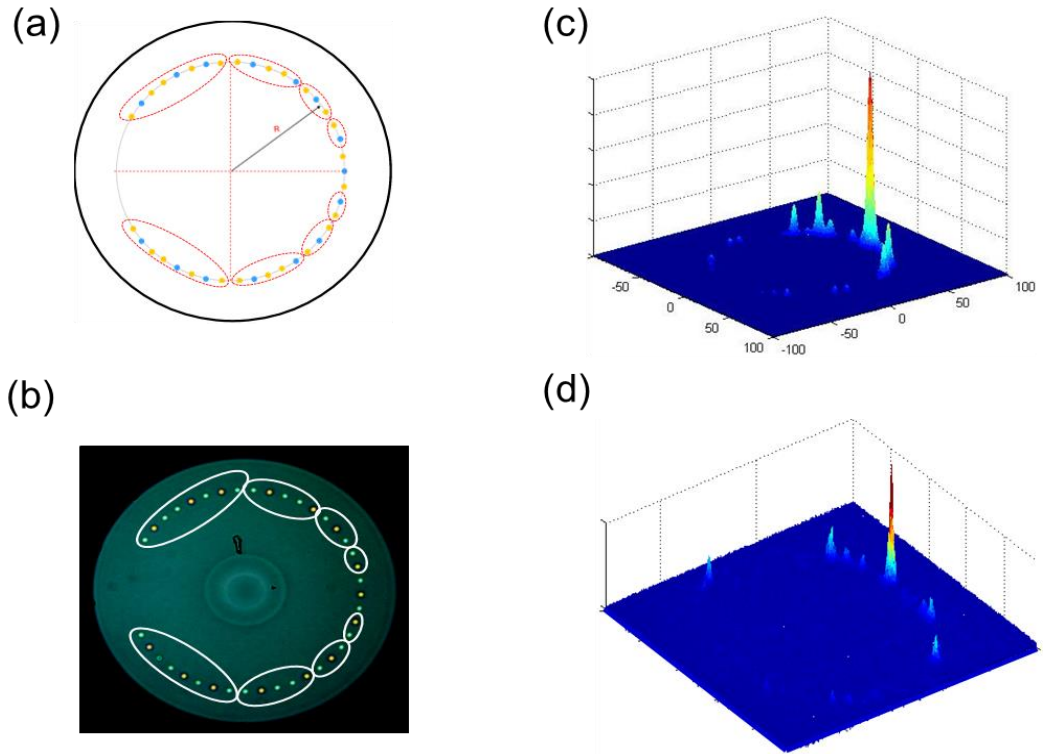

**Figure 6**. (a) Schematic of Fibonacci multicore ring fiber FMCRF of 39 cores that are regular-spacing in a ring of radius $R_1$=120 μm, cladding radius $R_2$ = 160 μm. There are two different SM waveguides *A* (blue) and *B* (orange) having difference $\Delta n_A = n_{coreA} - n_{clad}$ and $\Delta n_B = n_{coreB} - n_{clad}$, both cores *A* and *B* have same core diameter *a* = 4 μm, (b) Image of cross section of FMCRF fabricated with 39 cores, (c) calculated probability photon distribution of quantum walks in the FMCRF, and (d) Experimental data of the photon distribution. The experimental measurement of photon distribution at around 4.15cm of FMCRF and simulation results at ~ 4.15cm are in good agreement. Especially, strong localization at the center core are clearly shown in (c) and (d).

The simulation and experiment results in Figs. 5 and 6 show two typically different QWs in periodic (ordered) and quasiperiodic (deterministically disordered) core rings. In the ordered system - the MCRF, the expected distribution for typical QWs on a line has been observed experimentally with two strong ''ballistic'' lobes. On the hand, the quasiperiodic or deterministically disordered Fibonacci MCRF shows localized QWs as predicted by the simulations. We would like to stress here that, experimental results in both fibers can be further improved as several factors such as coupling misalignment, surface roughness, reflection by air/cladding interface can be reduced significantly from the current set up. Misalignment, roughness, air/cladding interfaces cause distortion and unwanted localization and interferences in the fiber. These issues can be resolved using index matching oil to reduce reflections at the interface between cladding and air. To achieve a short-length fiber with minimal roughness to the end faces from poor cleaving or from unwanted back-reflection from flat end faces, we fabricated a housing unit for the fiber made of angled-ferrules and/or canes filled completely or in part with index-matching oil and high-refractive index adhesive. The ferrules are polished at an angle or flat depending on tolerance for back-reflection. Optimization of the fiber-drawing process will improve the variation of core sizes in the fibers.

As can be seen from Figs. 5 and 6, LQWs in quasi-periodic FCRF are predictable and controllable in contrast with the ones in spatially random disordered structures. Furthermore, LQWs with the *symmetrical probability distribution* can be achievable in our new class of quasi-periodic photonics lattices. It is worth noting that Chandrashekar and Busch [22] have recently proposed to employ symmetrical LQWs for secure quantum memory applications. In order to achieve symmetrically distributed LQWs, they proposed to use temporally disordered operations in spatially ordered systems [22, 47]. However, their approach requires multiple quantum coins for temporally disordered operation which could be extremely difficult in practice.

At this point, we would like to note that localization in photonic lattices of waveguides constructed with the Fibonacci sequence *in distances* between nearest identical waveguides, both 1D and 2D, have recently been investigated theoretically and experimentally [48, 49]. Essentially, the Fibonacci arrays of waveguides considered in [48] are defined as the elements in Eq. 1 in this work; however, *A* and *B* were defined as the two fundamental distances 1 (unit) and $\tau$ between identical waveguides where $\tau = 1.618$ is the golden ratio of the Fibonacci sequence. Since the quasiperiodic properties of these element are determined by the Fibonacci sequences in the spacing of waveguides, they can be classified as quasiperiodic photonic lattices (QPLs) with the Fibonacci spacing sequence (FSS). It is important, however, to point out that the QPLs of identical waveguides constructed with the FSS described in [48, 49] have a quasiperiodic distribution of coupling coefficients (*off-diagonal* deterministic disorder) with all waveguides having the same propagation constants. Meanwhile our proposed Fibonacci MCRFs having cores constructed with a Fibonacci sequence with two different waveguides, resulting in both propagation constants and coupling coefficients are quasiperiodic or deterministically disordered distributions.

## 4. Conclusion

In conclusion, for the first time QWs and LQWs have been demonstrated in periodic and quasiperiodic multicore ring fibers, respectively. The new multicore-ring fibers (MCRF) provide a new scalable platform for experiments of QWs and other quantum effects that require extremely low insertion and large numbers of waveguides. The core structure in quasiperiodic FMCRF is constructed with Fibonacci sequence provides a new class of QPLs that possess both on- and off-diagonal deterministic disorders. Although the structures of the quasi-periodic Fibonacci lattices are straightforward to fabricate, the outcome results of LQWs are predictable and controllable, in contrast with LQWs in randomly disordered systems. The new class of QPL thus provide platforms for realizing LQWs and for investigation of quantum effects with both on- and off-diagonal deterministic disorders. Furthermore, the new proposed $j^{th}$-order QPLs are constructed as an orderly sequence chain of all elements up to $j^{th}$-order instead of individual $j^{th}$-order Fibonacci elements as in [48, 49], which are convenient to make QPLs symmetric that could be of benefit for some special applications [22, 47]. Finally, we would like to stress here that the rule of construction for symmetrical QPLs can be applied with other quasiperiodic sequences such as Thue-Morse [47, 50], and Rudin–Shapiro [47] sequences.

**Methods**

Fabrication of the multicore ring fibers. In order to demonstrate QWs and LQWs in periodic and quasi-periodic MCRFs, we have designed and fabricated this new type of fiber with ring of cores, both periodic and quasi-periodic ring of cores. The fabrication method of MCRFs and FMCRFs includes fabricating disc-shaped segments for cladding glass by polishing flats and drilling bores in the direction orthogonal to the flats stacking a number of disc-shaped segments to form a combined cladding glass part, so that bores of all the disks match, and inserting continuous core cane segments into the bores sealing flats of the disc segments to each other, sealing bores to the core canes, and drawing multi-core preform into a multi-core fiber. This method provides a robust and cost-effective process for manufacturing of precision multi-core fiber. Core-drilling provides precision and robustness. Stack-sealing enables the use of short and bulky precision drilled parts of cladding glass for constructing a multi-core fiber preform. Such preform may also include solid end pieces of glass for improved utilization of precision drilled parts in the fiber draw. Vacuum is used is used to hold the stack together in the furnace while it is being sealed. Sealing is achieved in all directions simultaneously. The use of temperatures above glass softening point enables sealing surfaces with just fine grind finish.

Modeling and simulation. As single photons QWs do not behave any differently from classically coherent wave propagation, and the distribution of light intensity corresponds to the probability distribution of photons. Therefore, we can use the beam propagation method (BPM), one of the most effective methods of light propagation simulation in complicated structures for simulating single photons QWs in our periodic and quasiperiodic MCRFs. The method of BPM has been well developed for decades and commercial software is also available. Details of the BPM method are presented generally in [43, 44], its applications for simulating QWs in photonics lattices [38, 39] and multicore fiber lasers in [45, 46].

**Author Contributions**

DTN proposed the idea of multicore ring fiber (MCRF) and quasi-periodic Fibonacci core ring fiber (FMCRF), designed the MCRF and FMCRF, performed modeling and simulation. TAN performed experiments and data analysis. RK fabricated the fibers. DAN and NFB discussed the idea of quasi-periodic multicore fibers, helped develop the methodology. All authors contributed to writing the paper.

**Competing financial Interests**

The authors declare no competing financial interests.